%% file: gauge.tex
\documentclass[12pt,aps,showpacs,amsmath,amssymb]{revtex4}

\usepackage{graphicx}
\usepackage{dcolumn}   
\usepackage{bm}        
\usepackage{color}
\usepackage{flafter}  
\newcommand{\domain}[1]{\cD(#1)}
\newcommand{\comment}[1]{[{\bf Com: }{\it #1}]}
\renewcommand{\comment}[1]{}

\let\vr\undefined

\include{my_commands}

\begin{document}
\title{Mixed gauge in strong laser-matter interaction}

\author{Vinay Pramod Majety, Alejandro Zielinski, and  Armin Scrinzi} 
\email{armin.scrinzi@lmu.de}
\affiliation{Physics Department, Ludwig Maximilians Universit\"at, D-80333 Munich, Germany}


\date{\today}

\begin{abstract}
We show that the description of laser-matter interaction in length gauge at short short
and in velocity gauge at longer distances allows for compact physical modeling 
in terms of field free states, rapidly convergent numerical approximation, and efficient
absorption of outgoing flux. The mathematical and numerical framework for using mixed 
gauge in practice is introduced. We calculate photo\-electron spectra generated 
by a laser field at wavelengths of 400$\sim$800\,nm from single-electron systems and
from the helium atom and hydrogen molecule. We assess the accuracy of 
coupled channels calculations by comparison to full two-electron solutions of the 
time-dependent Schr\"odinger equation and find substantial advantages of mixed 
over velocity and length gauges. 
\end{abstract}

\maketitle



\section{Introduction}

The choice of gauge in the interaction of strong, long wave-length fields with atoms and molecules
affects the physical modeling \cite{bauer05:gauge}, perturbative expansions as the $S$-matrix series 
\cite{faisal07:gauge,vanne09:gauge}, as well as the efficiency of numerical solutions \cite{cormier96:gauge}. 
For systems
where field quantization can be neglected and the field appears only as a time- and space-dependent 
external parameter, the wavefunctions in all gauges are unitarily related by time- and space-dependent 
multiplicative phases. An extensive discussion of gauge transformations in the context of
strong field phenomena can be found in \cite{bandrauk13:gauge}. When approximations are made,
the unitary equivalence of the wave-functions and the corresponding time-dependent Schr\"odinger equation (TDSE) 
is lost. An important example is the strong field approximation, 
where the system is assumed to either remain in the field-free initial state or move exclusively
under the influence of the laser field: the function representing the field-free initial state
depends on gauge. A similar situation arises, when a series expansion is 
truncated to a finite number of terms, as in an $S$-matrix expansion: the physical meaning of any
finite number of terms is different in different gauges. Also the discretization errors in 
a numerical calculation are gauge dependent. In particular, multiplication by 
a space-dependent phase changes the smoothness of the solution. As a result, numerical accuracy 
and convergence are depend on gauge. 

Mathematical and numerical aspects of using general gauges were adressed in 
Refs.~\cite{burke91:r-matrix,doerr92:r-matrix,robicheaux95:gauge}
in the context of Floquet theory and the time dependent Schr\"odinger equation, where various options
for mixing different gauges were discussed. In Refs.~\cite{burke91:r-matrix,doerr92:r-matrix}, mixing length, velocity, acceleration gauge
in the R-matrix Floquet method was achieved by the introduction of Bloch operators at the boundaries between the gauges. 
In Ref.~\cite{robicheaux95:gauge}, it was pointed out that alternatively the transition between regions can be taken to be 
differentiably smooth, which also allows application to the time-dependent Schr\"odinger equation (TDSE).

Here we will show that physical modeling on the one hand and efficient numerical
solution on the other hand impose conflicting requirements on the choice between the 
standard length and velocity gauges. We introduce the mathematical and numerical techniques
for resolving this conflict by using general gauges. We restrict our discussion to gauge transformations
in the strict sense, i.e.\ local phase multiplications, which does not include the acceleration ``gauge'',
as it involves a time-dependent coordinate transformation. Numerical performance of the various
gauges is compared on a one-dimensional model system. We show that, also with discontinuous transition 
between gauges, there is no need for the explicit inclusion of the $\delta$-like Bloch operators. 
We demonstrate validity and accuracy of mixed length and velocity 
gauge calculations in three dimensions by comparing to accurate velocity gauge results for the hydrogen atom
at 800\,nm wavelength. Finally, we combine local length gauge with asymptotic velocity gauge to compute photo\-electron 
spectra of $He$ and $H_2$ at a laser wavelength of 400~nm. Efficiency and accuracy of
the approach is shown by comparing to complete
numerical solutions of the two-electron problem. We find that mixed gauge allows low-dimensional approximations,
while in velocity gauge we achieve convergence only when we allow essentially complete two-electron dynamics.
We will conclude that few-body dynamics in the realm of bound states is more efficiently represented in length 
gauge, while the long-range representation of the solution prefers the velocity gauge.  

\section{Length, velocity, and general gauges}

In the interaction of small systems of sizes of $\lesssim 0.1\,nm$ with light at wavelength down to the 
extreme ultraviolet $\lambda\gtrsim 10\, nm$ one employs the dipole approximation, i.e.\ 
one neglects the variation of the field across the extension of the system
$\vEf(\vr,t)\approx \vEf(t)$. In length gauge, the interaction of a charge $q$ with the dipole field is
\beq
I_L(t)=q\vEf(t)\cdot\vr,
\eeq
while in velocity gauge it is
\beq\label{eq:inter-vel}
I_V(t)=-\vA(t)\cdot\vp+\frac12|\vA(t)|^2,\qquad \vA(t):=\int_{-\infty}^t q\vEf(\tau)d\tau.
\eeq
Here and below we use atomic units with $\hbar=1$, electron mass $m_e=1$, 
and electron charge $e=-1$, unless indicated otherwise.
In these two gauges the dependence of the 
dipole interaction operators on $\vr$ is particularly simple and
wavefunctions are unitarily related by
\beq
\Psi_V(\vr,t)=e^{i\vA(t)\cdot\vr}\Psi_L(\vr,t).
\eeq 
The transformation from length to velocity gauge is a special case of the general
gauge transformation, namely multiplication by a space- and time-dependent phase 
\beq
\Psi_g=U_g\Psi,\quad U_g:=e^{ig(\vr,t)}.
\eeq 
As $U_g$ is unitary, it leaves the system's dynamics unaffected, if 
operators and the time-derivative are transformed as
\beq
\mO\to O_g=U_g O U_g^*,\quad \ddt\to \ddt + U_g\dot{U}_g^*=\ddt - i\dot{g}.
\eeq
The above relations are valid for general $g(\vr,t)$ that are differentiable \wrt
$t$.  
If $g$ is twice differentiable in space, the gauge transforms of momentum operator and Laplacian 
are 
\beq
\vp=-i\vna\to \vp_g=-i\vna-\vB,\quad \Delta\to\Delta_g=[-i\vna-\vB]^2,\quad \vB:=\vna g.
\eeq
We see, in particular, that a gauge transform introduces a time- and space-dependent 
momentum boost $\vB(\vr,t)$.

A standard TDSE transforms as
\bea
\lefteqn{i\ddt\Psi=\left[\frac{\vp^2}{2} + V + q\vEf(t)\cdot\vr\right]\Psi}
\nonumber\\
\label{eq:tdse-gauge}
&\to&
i\ddt\Psi_g=\left[ \frac{(\vp-\vB)\cdot(\vp-\vB)}{2} + V + q\vEf(t)\cdot\vr-\dot{g}\right]\Psi_g(\vr,t).
\eea
By explicitly writing the dot-product in the kinetic energy we emphasize that $\vp$ does not
commute with space-dependent $\vB(\vr,t)$ and space derivatives of $\vB$ appear in the Hamiltonian. 
The velocity gauge interaction Eq.~(\ref{eq:inter-vel}) requires spatially 
uniform $\vB(\vr,t)\equiv \vA(t)$. More generally, any time-dependence 
of the potential energy $V(\vr,t)$ can be transformed into 
a time- and space-dependent momentum by defining
\beq
g_V(\vr,t)=\int^t V(\vr,t') dt'.
\eeq

\subsection{Discontinuous gauge transformation}
The local phase multiplication need not be continuously differentiable or even continuous in space.
One only must make sure that the gauge tranformed differential operators $\vna_g$ are defined 
on functions $\chi$ from a suitable domain $\domain{\vna_g}$.
With discontinuous $g$, formally, $\delta$-like operators appear in
Eq.~(\ref{eq:tdse-gauge}). $\domain{\vna_g}$ must be adjusted to compensate for those terms. 
The very simple, mathematically correct solution is to choose $\domain{\vna_g}=U_g\domain{\vna}$, i.e. 
functions of the form
\beq\label{eq:domain}
\chi(\vr,t)=U_g(\vr,t)\varphi(\vr),\quad\varphi\in\domain{\vna},
\eeq
where the $\varphi(\vr)\in\domain{\vna}$ are differentiable.

With $g(\vr,t)=0$ at ranges $|\vr|<R_g$ and $g(\vr,t)\to\vA(t)\cdot\vr$ for large $|\vr|$, one can switch from lenght to 
mixed gauge. Clearly, the particular form of the transition and the corresponding
modulations of the wavefunction do not bear any physical meaning.
Still, accurate modeling of the transition is needed to correctly connect
the  length to the velocity gauge part of the solution. In the transition region one needs
to densly sample the solution, which increase the number of expansion coefficients. 
In many cases, this will also increase the stiffness of the time propagation equations and 
further raise the penalty for 
a smooth transition.

When $g$ or any of its derivates is discontinuous, the discretization in the vicinity of the
discontinuity must be adjusted appropriately. Because of the lack of 
differentiability, any higher order finite difference scheme or approximations by analytic basis
functions will fail to improve the approximation or may even lead to artefacts. The general solution for
this problem is to explicitly build the known non-analytic behavior of the solution into the discretization.
With the finite element basis set used below, this is particularly simple, as well-defined discontinuities
can be imposed easily. We will also demonstrate below that a spatially abrupt transition does not increase stiffness
and allows calculations with similar efficiency as in uniform velocity gauge.

\subsection{Gauge in the strong field approximation (SFA)}

When we describe a physical process in terms of a few quantum mechanical states it
is implied that the system does not essentially evolve beyond those states.
The ``strong field approximation'' (SFA) is a simple model of this kind, which plays
a prominent role in strong field physics. 
One assumes that an electron either resides in its initial state or, 
after ionization, moves as a free particle in the field, whose effect largely exceeds the atomic
binding forces. 

The SFA must be reformulated appropriately depending on the gauge on chooses.
Let $\Phi_0(\vr)$ be the initial state in absence of the field. The physical picture above implies
that the velocity distribution of the initial state remains essentially unchanged 
also in presence of the field. However, using the same function $\Phi_0$ for all gauges,
effectively leads to a set of different models with different, time-dependent velocity distributions
for different gauges. In length gauge the operator $-\frac{i}{m_e}\vna$ has the meaning of a velocity of
the electron and the velocity distribution is independent of the field:
\beq
n_L(\vp,t)\equiv n_0(\vp)=|\tilde{\Phi}_0(\vp)|^2.
\eeq 
In contrast, in velocity gauge, the velocity distribution varies with time as
\beq
n_V(\vp,t)=\frac{1}{m_e}|\tilde{\Phi}_0[\vp+\vA(t)]|^2.
\eeq 
The difference becomes noticeable when the variation of $\vA(t)$ is not negligible compared to the 
width of the momentum distribution. This is typically the case in strong field phenomena.
Findings that SFA in length gauge better approximates the 
exact solution in cases where the picture remains suitable at all \cite{Awasthi2008,busuladzic10:gauge}
are consistent with this reasoning. 

\subsection{Single active electron (SAE) approximation}

The gauge-dependent meaning of eigenstates has important consequences for the numerical approximation
of few-electron systems. The functions corresponding to few-electron bound states have their intended
physical meaning only in length gauge. In velocity gauge, the same functions correspond to time-varying
velocity distributions. The problem affects the ``single active electron'' (SAE) approximation, where one lets
one ``active'' electrons freely react to the laser field, but freezes all other electrons in their 
field-free states. Below we will demonstrate that this ansatz generates artefacts in velocity gauge.

As a simple illustration of the problem let us consider two non-interacting electrons
with the Hamiltonian $H(x,y)=h(x)+h(y)$. The ansatz for the solution 
$\Phi(x,y,t)=\varphi(x,t)\chi(y,t)-\chi(x,t)\varphi(y,t)$ is exact, if the functions $\varphi$ and $\chi$ are unrestricted.
Matrix elements of the Hamiltonian are
\bea
\l \Phi | H | \Phi\r& =& \l \varphi | h |\varphi\r\l \chi | \chi\r +  \l \varphi |\varphi\r\l \chi|h | \chi\r\\
&&-  \l \varphi | h |\chi\r \l \chi|\varphi\r - \l \varphi |\chi\r \l \chi|h|\varphi\r.
\eea
In an exact calculation, the exchange terms in the second line vanish, 
if $\l\varphi|\chi\r=0$ initially, as the unitary  time evolution maintains orthogonality. 
However, if we restrict the time evolution of one of the functions, say $\varphi$, orthogonality is violated and 
unphysical exchange terms appear in the Hamiltonian matrix as the system evolves. 
Their size depends on the extend to which orthogonality is lost.
If e.g. $\varphi$ remains very close to its field-free state (e.g. if it is closely bound), then in length gauge
the time evolution is well approximated as $\varphi(t)\approx \varphi(0)$. However, depending on the size of $\vA(t)$, 
in velocity gauge this does not hold and the exchange terms become sizable. 

With interacting electrons, the direct term (Hartree potential) 
of electron-electron interactions is unaffected, as it only depends on the gauge-invariant electron 
density. In the exchange terms, however, the frozen orbitals with their length gauge meaning are inconsistently 
combined with the velocity gauge functions of the active electron. 

The same gauge dependence appears also when the non-active electrons are not frozen in their initial states, 
but restricted in their freedom to evolve. We will demonstrate the superiority of length gauge 
for $He$ and $H_2$ with limited freedom for the non-active electron.

\subsection{Gauge in numerical solutions}

While length gauge lends itself to intuitive interpretation and modeling, velocity gauge performs
better in numerical calculations \cite{cormier96:gauge}. Fewer
discretization coefficients can be used and the stiffness of the equations is reduced.
This is due to the dynamics of free electrons in the field.
From Eq.~(\ref{eq:tdse-gauge}) one sees that for a free electron ($V_0=0$) the velocity gauge canonical
momentum $\vp=-i\vna$ is conserved. In contrast, in length gauge, momenta are boosted by $\vA(t)$, 
reflecting the actual acceleration of the electron in the field. As large momenta correspond to short 
range modulations of the solution, length gauge requires finer spatial resolution than velocity
gauge. This modulation affects numerical efficiency, when the variation of $\vA(t)$ is comparable or exceeds the 
momenta occurring in the field-free system. We will illustrate this below on one- and three-dimensional examples.

A second important reason for velocity gauge in numerical simulation is the use of infinite range 
exterior complex scaling (irECS) \cite{tao12:ecs-spectra} for absorption at the box boundaries. This method is highly 
efficient and free of artefacts, but it cannot be applied for systems with length gauge asymptotics, as clearly observed 
in simulations \cite{scrinzi10:irecs}. An intuitive explanation of this fact can be found in \cite{mccurdy04:complex_scaling}
and the wider mathematical background is laid out in \cite{reed-simon82}. 

\subsection{Mixed gauge}

The conflicting requirements on gauge can be resolved by observing that bound states are, by definition,
confined to moderate distances, whereas the effect of phase modulation is important for free electrons,
usually far from the bound states. 
Using length gauge within the reach of bound states and velocity gauge otherwise largely unites
the advantage of both gauges: locally, the system can be modeled intuitively, while at the same time
maintaining efficient numerical spatial discretization and asymptotics suitable for absorption.

\section{Implementations and examples}

\subsection{TDSE in one dimension}
We use a basic model for discussing the various options for implementing mixed gauges.
We solve the TDSE with the  length gauge Hamiltonian 
\beq\label{eq:1d-length}
H_L(t)=-\frac12\ddxx - \frac{1}{\sqrt{x^2+2}}-x\Ef(t).
\eeq
In absence of the field, the ground state energy is exactly $-0.5$.

Spatial discretization is by a high order 
finite element basis, which is described in detail in Ref.~\cite{scrinzi10:irecs}. 
Apart from being numerically robust, the basis is flexible, 
which allows, in particular, easy implementation of the discontinuity Eq.~(\ref{eq:domain}).

In general, basis functions $|j\r$ used for spatially discretizing the TDSE do not need to be twice 
differentiable. Rather, they can have discontinuous first derivatives.
Although the second derivative is not defined as an operator on the Hilbert space, 
matrix elements can be calculated correctly by using the symmetrized 
form
\beq
\l k | -\ddxx | j\r :=\l \ddx k| \ddx j\r.
\eeq 
This relaxed condition on the differentiability is explicitly used with finite element bases, where 
usually first derivatives are discontinuous at the boundaries between elements.

The ansatz
\beq
|\Psi(x,t)\r\approx \sum_{j=1}^N |j\r c_j(t)
\eeq 
leads to the system of ordinary differential equations for the expansion coefficients $\vc$, $(\vc)_j=c_j$
\beq\label{eq:ode}
i\ddt \mS \vc(t) = \mH(t) \vc(t)
\eeq
with the matrices
\beq
\mH_{kj}(t)=\l k| H(t) |j\r,\quad \mS_{kj}=\l k | j \r.
\eeq
For time-integration, we use the classical 4th order explicit Runge-Kutta solver. As an 
explicit method it is easy to apply, but it is also susceptible to the stiffness
of the system of equations (\ref{eq:ode}). This is a realistic setting, as in many
practical implementations explicit time-integrators are used. For the present purpose, 
it clearly exposes the numerical properties of the different gauges.
In the one-dimensional case we use a simulation box large enough such that reflections
at the boundary remain well below the error level.

\subsubsection{Mixed gauge implementations (1d)}

Matrix elements of the kinetic energy are always computed in the explicitly hermitian form
\beq
\l k | [-i\ddx - B]^2 | j\r =  \l \ddx k | \ddx  j\r - i\l \ddx k| B j\r +  i\l B k| \ddx j\r + \l k | B^2 | j\r,
\eeq
which also avoids the calculation of spatial derivatives of $B(x,t)$. Also, for non-differentiable $B$,
no $\delta$-like operators appear.

For differentiable functions $B(x,t)$ we have domains $\domain{\Delta_g}=\domain{\Delta}$ and no adjustments need
to be made for the basis functions. To avoid loss of numerical approximation order, one must make
sure that $B(x,t)$ is smooth to the same derivative order as the the numerical approximation. 
We will refer to this case as ``smooth switching''.

With a discontinuous change of gauge
\beq
U_D=\left\{\bar 1 &\text{ for }|x|<R_g\\ e^{iA(t)x} & \text{ for }|x|>R_g\ear\right. 
\eeq
the Hamiltonian is
\bea\label{eq:Hdiscontinuous}
H_D(t)&=&\left\{\bar
H_L(t)&\text{ for }|x|<R_g\\
H_V(t)&\text{ for } |x|>R_g,\ear\right.
\eea
where we denote the standard velocity gauge Hamiltonian as $H_V(t)$.
There appear time-dependent discontinuities at the ``gauge radius'' $R_g$
\beq
\Psi_D(\pm R_g+\ep)=e^{ iA(t)R_g}\Psi_D(\pm R_g-\ep).
\eeq
In a finite element basis such discontinuities can be imposed explicitly.
Among others it leads to a time-dependent overlap matrix, whose inverse at each time-step 
can be obtained at low computational cost by low-rank updates. 
The technical details on this will be presented elsewhere.  

One can avoid discontinuities at $R_g$ by ``continuous gauge switching''
\beq
U_C=\left\{\bar 1 &\text{ for }|x|<R_g\\ e^{iA(t)(x\mp R_g)} & \text{ for } x\gtrless\pm R_g\ear\right. 
\eeq
with the Hamiltonian
\bea\label{eq:Hcontinuous}
H_C(t)&=&\left\{\bar
H_L(t)&\text{ for }|x|<R_g\\
H_V(t)\pm q\Ef(t)R_g&\text{ for } x\gtrless \pm R_g.\ear\right. 
\eea
Note that $U_C$ is continuous, but not differentiable at $R_g$, which leads to discontinuous first derivatives
in the solution. As discussed above, a 
finite element basis admits discontinuities of the derivatives and there is no
penalty in the numerical approximation order, if one ensures that $x=\pm R_g$  fall onto
element boundaries. 

The $U_D$ and $U_C$ formulations are nearly equivalent in their numerical behavior, as 
\beq
U_C = U_0U_D,\quad U_0=\left\{\bar 1 &\text{ for }|x|<R_g\\ 
                   e^{\mp iA(t) R_g} & \text{ for } x\gtrless \pm R_g.\ear\right.
\eeq
The respective solutions differ only by the phases  $e^{\mp iA(t) R_g}$: 
\beq
\Psi_C=U_0\Psi_D.
\eeq
Depending on $R_0$, the time-dependence of this phase is slow compared to phase-oscillations due to high 
energy content of the solution and does not change stiffness for numerical integration.

In the following, we compare mixed gauge in the three forms given here with pure 
length and velocity gauge calculations.

\subsubsection{Comparison of the gauges}
We compare electron densities $n(x)$ at the end of the laser pulse. 
Size of the spatial discretization and the number of time-steps are adjusted to reach the same 
local error $\ep(x)$ in all gauges relative to a fully converged density $n_0(x)$. 
For suppressing spurious spikes at near-zeros of the density, we include some averaging into the
definition of the error:
\beq\label{eq:smoothed-error}
\ep(x)=2\Delta x |n_0(x)-n(x)|/\int_{-\Delta x}^{\Delta x} dx' n_0(x')
\eeq
with $\Delta x=1$.

\begin{figure}
  \includegraphics[width=0.9\linewidth]{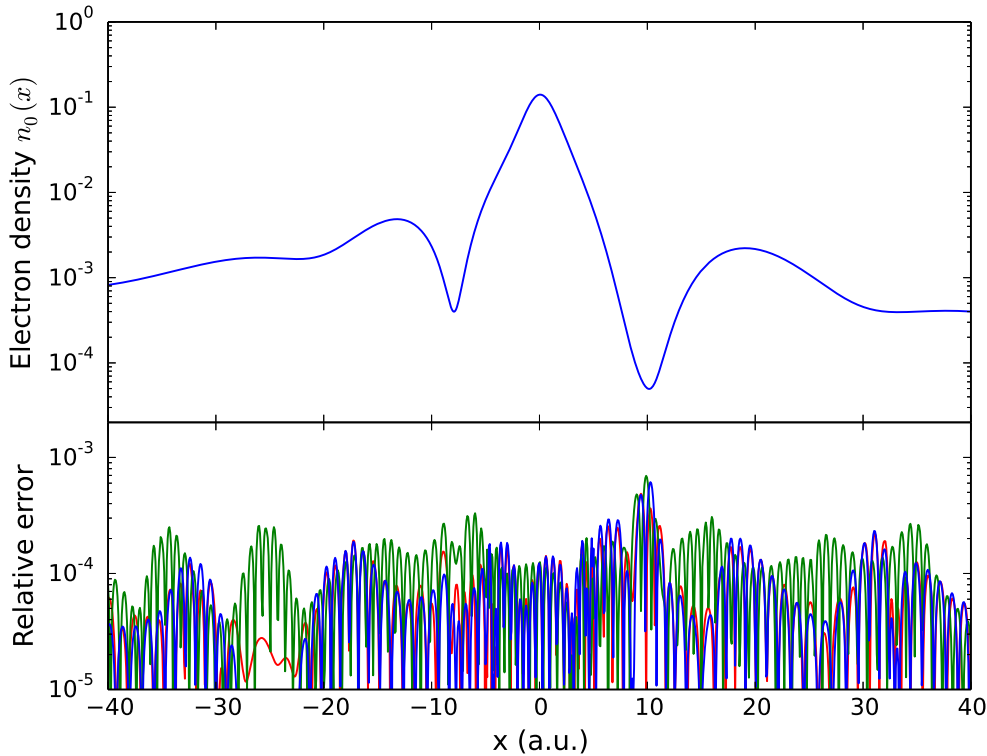}
\caption{\label{fig:1d-grids}
Electron density of the one-dimensional model system at the end of a single-cycle pulse (see text for exact 
pulse definition.
Upper panel: fully converged velocity gauge calculation with simulation box size [-1000,1000], 
finite element order 20, $N\approx5500$ discretization coefficients, $T=12\times10^4$ time steps.
Lower panel: Relative errors Eq.~(\ref{eq:smoothed-error}) in various gauges. Velocity gauge, 
$N\approx3000$, T=$7.1\times10^4$ (red line), length gauge,
$N\approx4000$,T=$8.8\times10^4$ (green), mixed gauge, $R_g=5$, $N\approx3000$,T=$7.1\times10^4$ (blue). Errors of velocity and mixed gauge nearly coincide.
}
\end{figure}

We use a single cycle 800 nm pulse with  $\cos^2$-shape 
and peak intensity $2\times10^{14}\,W/cm^2$, which leads to about 25\% ionization of this 1-d system. 
The $x$-axis is confined to [-1000,1000] with Dirichlet
boundary conditions, discretized by  finite elements of polynomial order 20.    
Figure~\ref{fig:1d-grids} shows results in the different gauges. 
Velocity gauge requires $N\approx3000$ linear coefficients (150 elements) 
and about $T\approx71000$ time steps for accuracy  $\ep(x)\lesssim 10\inv4$.
Length gauge has the largest discretization with $N\approx4000$ and $T\approx88000$, 
amounting to an overall increase in computation time of almost a factor 2. The larger number of time-steps
arises because the explicit propagation scheme is sensitive to the stiffness of the equations, which can grow $\propto N^2$.
The actual increase of times steps does not exactly reflect this behavior, as a finer discretization is used in the 
inner region to obtain comparably accurate initial states in all calculations. Stiffness from this discretization is always
present in the calculations.
For the mixed gauge, we use continuous switching Eq.~(\ref{eq:Hcontinuous}) with $R_g=5$. With this small length gauge section,
the same discretization as in the velocity gauge can be used with $N\approx 3000$ and $T\approx 71000$.

We also investigated the effect of a smooth transition between the gauge over an interval of size $S$, using 
\beq
g(x,t)=\left\{\bar
0 & \text{ for } |x|<R_g\\
xs(x)A(t) & \text{ for } |x|\in[R_g,R_g+S]\\
xA(t) & \text{ for } |x|>R_g+S
\ear\right.
\eeq
where $s(x)$ is a 3rd order polynomial smoothly connecting the length with the velocity gauge region. 

For a smoothing interval $S=5$, we need a rather dense discretization by 18th order polynomials on the small interval
to maintain the spatial discretization error of $\approx 10\inv4$. While this leads only to a minor increase
in the total number of discretization coefficients, it significantly increases the stiffness of the equations 
requiring $T=1.4\times10^5$ time steps. With smoothing $S=10$, stiffness is reduced and $T\approx10^5$, which 
still exceeds by 50\% the number of time steps with continuous, but non-differentiable transition. 

The dependence on $S$ is not surprising: the correction terms to the kinetic
energy involve derivatives of $s(x)$, which grow inversely proportional to the size of the transition region,
leading to large matrix elements. Thinking in terms of the solution, we need to follow a rather strong 
change in temporal and spatial behavior of the solution, which necessitates the dense grid. With the sudden transition,
this change is reduced to a single discontinuity, whose behavior we know analytically. It can either be build explictly 
into the solution,
when using the discontinuous Hamiltonian $H_D(t)$, Eq.~(\ref{eq:Hdiscontinuous}), 
or be left to be adjusted numerically with the continuous Hamiltonian $H_C(t)$,  Eq.~(\ref{eq:Hcontinuous}). 
We conclude that, wherever technically possible, a sudden transition is to be preferred.


\subsection{Mixed gauge for the Hydrogen atom}

The length gauge Hamiltonian for the hydrogen atom in a laser field is
\beq
H_L(t)=-\frac12\Delta -\frac1r - \vEf(t)\cdot\vr.
\eeq
The velocity gauge Hamiltonian is
\beq
H_V(t)=-\frac12[-i\vna-\vA(t)]^2 -\frac1r.
\eeq
In three dimensions, problem size grows rapidly and truncation of the simulation volume is advisable. 
As the error free absorbing boundary method irECS \cite{tao12:ecs-spectra} is incompatible with length gauge calculations, 
in this section we only compare velocity to mixed gauge calculations. Following the findings of the 
one dimensional calculation, we use continuous gauge switching for its numerically efficiency and moderate programming effort. 
In three dimensions, it can be defined as
\beq
U_C=\left\{\bar 
1 &\text{ for }r<R_g\\
\exp\left[i\vA(t)\cdot\ur(r-R_g)\right]&\text{ for } r>R_g,  
\ear\right.
\eeq
denoting $\ur:=\vr/r$.
The corresponding Hamiltonian is
\beq
H_C(t)=\left\{\bar
H_L(t)&\text{ for } r<R_g\\
\left[-i\vna-\vB(\vr,t)\right]^2-\frac{1}{r}-\vEf(t)\cdot\ur R_g &\text{ for } r>R_g.
\ear\right.
\eeq
The gradient of the angle-dependent phase introduces an extra quadrupole type coupling:
\beq
\vB(\vr,t)=\vna\left[\vA(t)\cdot\ur(r-R_g)\right]=\vA(t)\left[1-\frac{R_g}{r}\right]+\left[\vA(t)\cdot\ur\right]\frac{\ur R_g}{r}.
\eeq
$H_C(t)$  asymptotically coincides with standard velocity gauge as $|\vr|$ tends to $\infty$. 
In an expansion into spherical harmonics, the quadrupole terms introduce additional non-zeros 
into the Hamiltonian matrix, which increase the operations count for applying the Hamiltonian by 
$\sim$60\%.

\subsubsection{Comparisons}
For the numerical solution we use polar coordinates with a finite element basis on the radial coordinate
and spherical harmonics for the angular dependence. A discussion of the basis can be found
in \cite{tao12:ecs-spectra}. We assume linear polarization and fix the magnetic 
quantum number at $m\equiv 0$. 
We use a $\cos^2$-shaped pulse with 3 optical cycles FWHM at central wavelength $\lambda=800\,nm$
and and peak intensity $2\times10^{14}\,W/cm^2$, which leads to about 16\% ionization. 

We compare the errors of the different gauges in the angle-integrated electron density $n(r)$ at the 
end of the pulse and in the photo\-electron spectra. The spectra are computed by the tSURFF method described in 
Refs.~\cite{tao12:ecs-spectra,scrinzi12:tsurff}. Errors are again defined relative to a fully converged velocity gauge
calculation.

On the radial coordinate we use 5 finite elements of order 16 up to 
radius $R_0=25$ in all gauges. Beyond that, the solution is absorbed by irECS.
The stronger phase oscillations of the length gauge solution requires more 
angular momenta compared to velocity gauge \cite{cormier96:gauge}. 

\begin{figure}
 \includegraphics[width=0.9\linewidth]{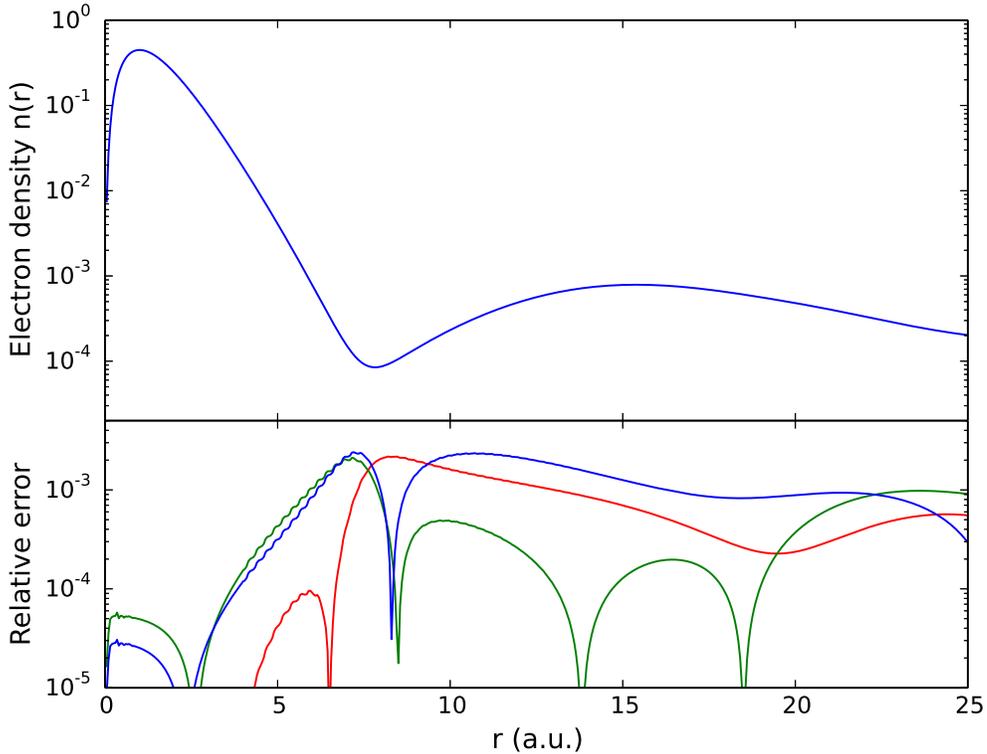}
\caption{\label{fig:3d-density}
Velocity vs. mixed gauge for the hydrogen atom in three dimensions. 
Upper panel: electron-density up to the absorption radius $R_0=25$, fully converged calculation.
Lower panel: Relative errors  Eq.~(\ref{eq:smoothed-error}) compared with a fully converged calculation.
Red: velocity gauge $L_{max}=21$,
green: mixed gauge, $R_g=5, L_{max}=30$, 
blue: mixed gauge at  $R_g=20, L_{max}=35$. Radial discretization by $N=80$ functions.
}
\end{figure}

Figure~\ref{fig:3d-density} shows the relative errors in $n(r)$ of a velocity gauge calculation 
with $L_{max}=22$ angular momenta with mixed gauge calculations at two different gauge radii 
$R_g=5,L_{max}=30$, and $R_g=20,L_{max}=35$. As expected, the mixed gauge calculation needs
 higher $L_{max}$ as $R_g$ increases.    

\begin{figure}
 \includegraphics[width=0.9\linewidth]{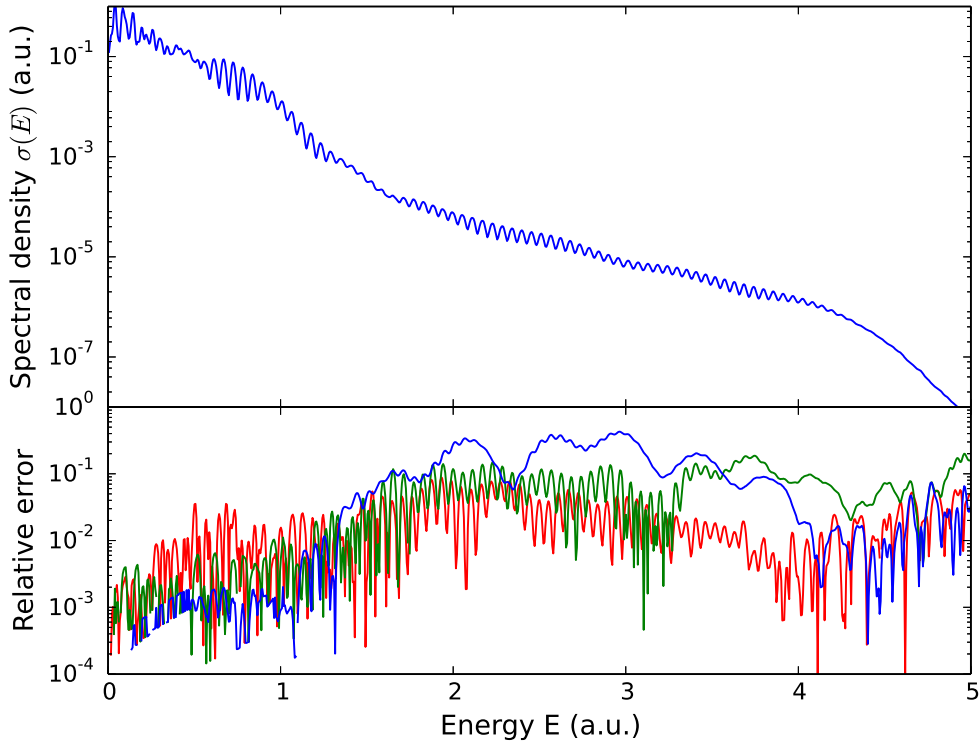}
\caption{\label{fig:3d-spectra}
Photoelectron spectrum (upper panel). For pulse parameters see text. Lower panel: relative errors. 
Red: velocity gauge $L_{max}=21$,
green: mixed gauge, $R_g=5, L_{max}=30$, 
blue: mixed gauge at  $R_g=20, L_{max}=40$. Radial discretization as in Fig.~\ref{fig:3d-density}.
}
\end{figure}

The same general error behavior of the different gauges is also found in the photo\-electron
spectra, Figure~\ref{fig:3d-spectra}. Here, the $R_g=20$ requires even more angular momenta $L_{max}=40$. This may be due
to the particular sensitivity of photo\-electron spectra to the wavefunction at the radius 
where the surface flux is
picked up and integrated, in the present case at $r=25$.

\subsection{Helium atom and $H_2$ molecule}
The length gauge Hamiltonian for a two-electron problem interacting with a dipole laser field is
\beq
H_L(t)=\sum_{k=1,2}\left[-\frac12\Delta_k - \frac{1}{|\vr_k-\vR/2|} - \frac{1}{|\vr_k+\vR/2|} 
- \vEf(t)\cdot\vr_k\right]+\frac{1}{|\vr_1-\vr_2|}.
\eeq
This includes the $H_2$ molecule at fixed internuclear distance $\vR=(0,0,1.4 au)$ and the helium atom $|\vR|=0$.
We assume linear polarization in $z$-direction.

We compare total photo\-electron spectra.
As a reference, we solved the two-electron (2e) TDSE 
fully numerically in velocity gauge using a single-center expansion. Details of this calculation will be 
reported elsewhere \cite{zielinski:solver}. 
Photo\-electron spectra for the various ionic channels were computed using the 2e form of tSURFF 
(see \cite{scrinzi12:tsurff}).
As 2e calculations are very challenging at long wavelength, we use a 3-cycle pulse at somewhat shorter wavelength
of $\la=400\, nm$ and an intensity of only $1\times 10^{14}W/cm^2$.
To facilitate the extraction of photo\-electron momenta, all potentials where 
smoothly turned off beyond distances $|\vr_i|>25\ a.u.$ as described in Ref.~\cite{scrinzi12:tsurff}.

We compare the 2-electron calculation with a coupled channels computation using the expansion
\beq
|\Psi(\vr_1,\vr_2,t)\r=|0\r c_0(t) + \sum_{I,j}\cA\left[|I\r|j\r\right] c_{Ij}(t),
\eeq  
which includes the field-free neutral ground state $|0\r$ and the ionic states $|I\r$ multiplied by the same
single-electron basis functions $|j\r$ as for the hydrogen atom. Anti-symmetrization is indicated by $\cA[\ldots]$.
The neutral ground state $|0\r$ as well as the ionic states $|I\r$ were obtained 
from the COLUMBUS quantum chemistry package \cite{lischka11:columbus}. Calculations were performed in velocity and 
mixed gauge (continuous switching), as described for the hydrogen atom. Details of the coupled channels method
will be reported reported elsewhere \cite{majety:solver}.

\begin{figure}
\includegraphics[width=0.9\linewidth]{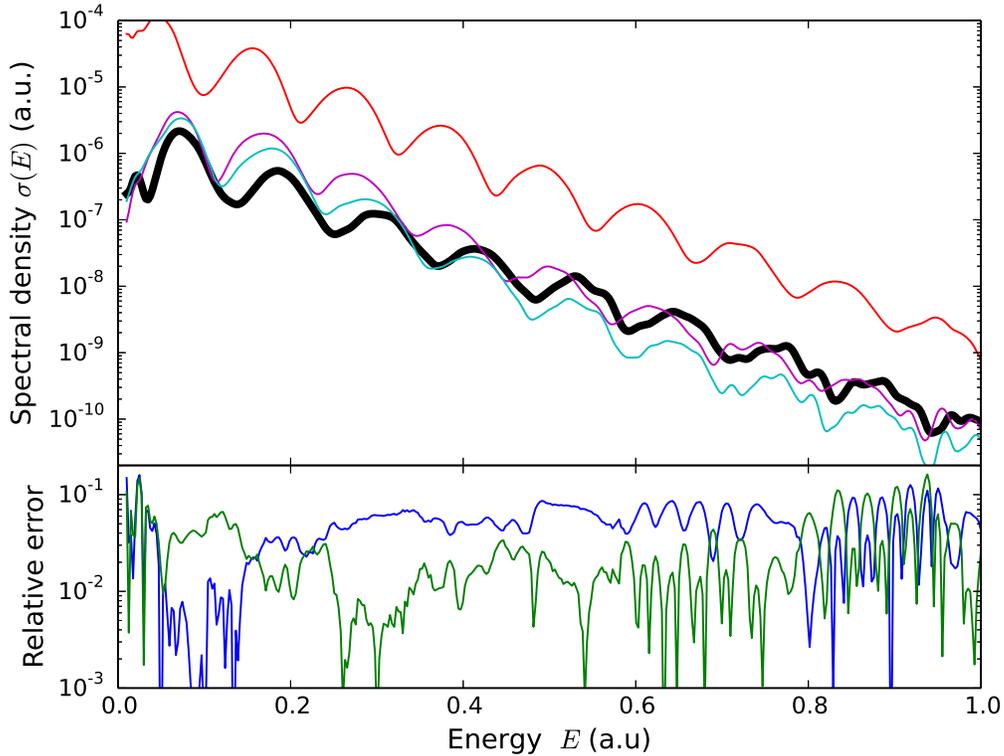}
\caption{\label{fig:he-spectra} Photo\-electron spectrum of helium at 400 nm. Upper panel, thick line: 
full 2e calculation in velocity gauge, thin lines coupled channels plus the neutral ground state:
include n=1 ionic state (red), $n\leq2$ ionic states (magenta),  $n\leq3$ s and p ionic states (cyan). All mixed gauge
coupled channels calculations nearly coincide with the 2e calculation.
Lower panel: relative difference between mixed gauge and the 2e calculation. Include only ionic ground n=1 state (blue)
and all $n\leq2$ states (green). \comment{why is n=5 worse than n=1 at low energies?}
}
\end{figure}

\begin{figure}
\includegraphics[width=0.9\linewidth]{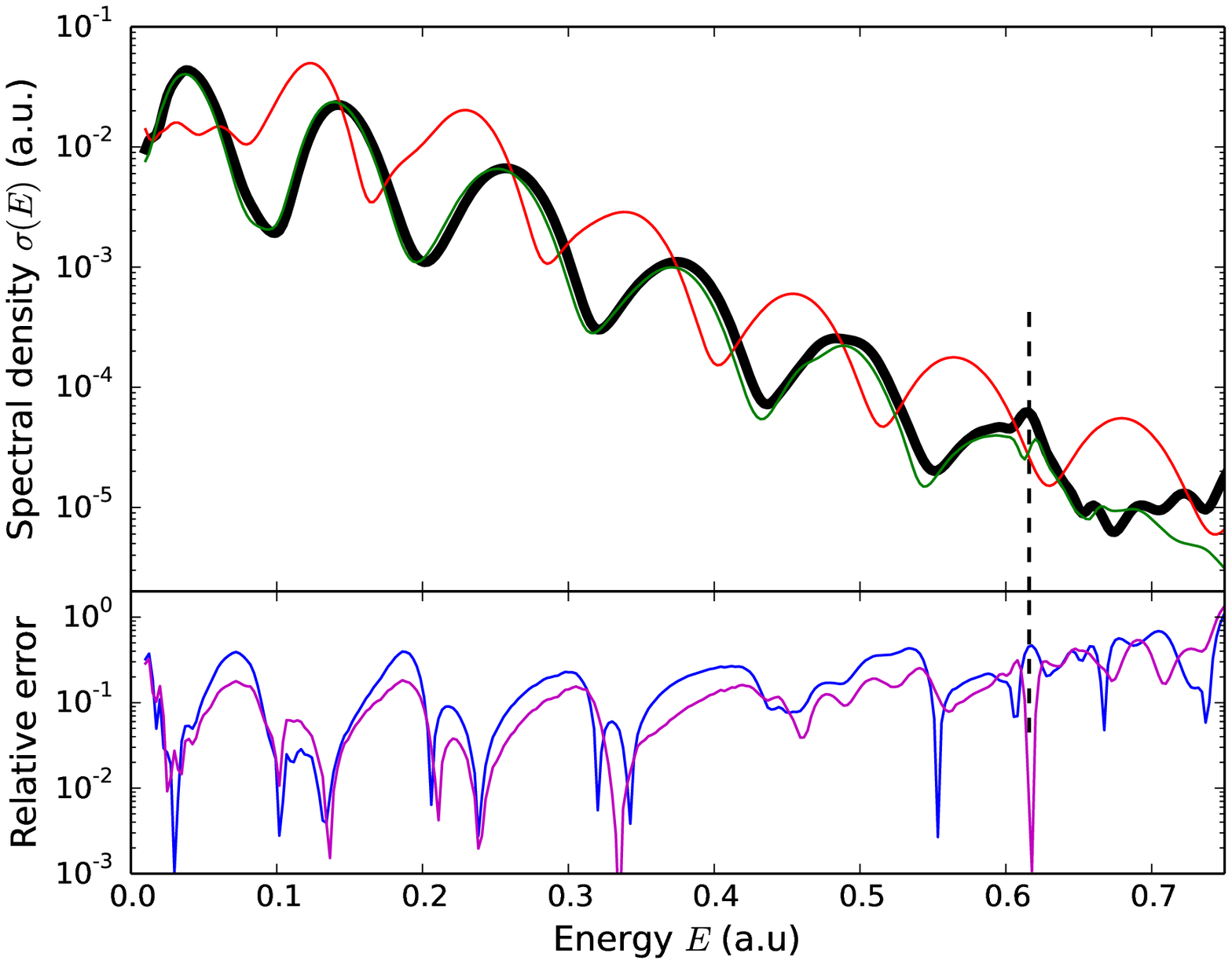}
\caption{\label{fig:h2-spectra}
 Photo\-electron spectrum of $H_2$ at 400 nm. Upper panel, thick line: 
full 2e calculation in velocity gauge, thin lines coupled channels plus the neutral ground state:
include only the $\sigma_g$ ionic ground state (red), include the 6 lowest $\sigma$ and $\pi$ ionic states (green). 
Lower panel: relative difference between mixed gauge and the 2e calculation. Include only $\sigma_g$ ionic ground state (blue)
and include the 6 lowest $\sigma$ and $\pi$ ionic states (magenta). The dashed line marks the resonance position.
}
\end{figure}

By the arguments above, in the coupled channels basis we expect the mixed-gauge calculation to converge better than the 
velocity gauge calculation: the COLUMBUS wavefunctions $|0\r$ and $|I\r$ have their intended physical meaning
only in length gauge. Figures~\ref{fig:he-spectra} and \ref{fig:h2-spectra} confirm this expectation.

For helium, Figure~\ref{fig:he-spectra}, the velocity gauge 2e calculation agrees well with the mixed gauge
coupled channels calculation using the neutral and only the 1s ionic state. 
With the 5 ionic states with principal quantum number $n\leq2$ the error is $\sim2\%$ for a large part of the spectrum up to 1 au. 
In contrast, the
single-ion velocity gauge calculation is far off. It does somewhat approach the 2e result when the number of 
ionic states is increased to include the ionic states up to $n=3$. In velocity gauge, convergence could not be achieved 
for two reasons. One reason is a technical limitation of the coupled channels code, which uses Gaussian basis functions that do 
not properly represent the higher ionic states.
The second reason is more fundamental: in velocity gauge the ionic cores transiently contain significant
continuum contributions, which are not included in our ionic basis by construction.

The error pattern is similar in $H_2$, but the accuracy of all calculations is poorer,  Fig.~\ref{fig:h2-spectra}: 
2e and coupled channel mixed gauge calculations qualitatively
agree already when only the $\sigma_g$ ionic ground state is included. 
With the lowest 6 ionic $\pi$ and $\sigma$ states the two spectra differ by $\lesssim 20\%$. Remarkably, the height of
a small resonant peak at $\sim 0.62\,au$ is faithfully reproduced in mixed gauge with 6 ionic states.
 The resonance can be tentatively assigned to the near degerate second and third
$^1\Sigma_u^+$ doubly excited states of $H_2$ at the internuclear equlibrium distance of 1.4 au
(see Ref.~\cite{sanchez97:h2}).  \comment{Check for consitency}
As a note of caution, the single center expansion used in the 2e code converges only 
slowly for $H_2$ and cannot be taken as an absolute reference. 
The coupled channels velocity gauge calculation is off by almost two orders of magnitude when only a single ionic state 
is included.
With 6 ionic states it compares to the full 2e on a similar level as the mixed gauge. However, in velocity
gauge the resonance is not reproduced correctly. 

For both systems, analogous results were found at shorter wavelength down to $\la=200\,nm$. At even shorter wavelength and 
realistic laser intensities, gauge questions are less important as the magnitude of $|\vA(t)|\propto \la$.

\section{Conclusions}
In summary, we have shown that a transition between gauges within the same calculation bears substantial advantages 
and requires only moderate implementation effort. For low-dimensional problems, the advantage can be technical, such as 
reducing the 
size of the spatial discretization and the equations' stiffness. We have shown that with a suitably chosen 
basis a sudden, non-differentiable transition from length to velocity gauge is preferable  
over a differentiably smooth transition in terms of both, simplicity of implementation and numerical efficiency. 

Mixed gauge opens the route to a highly efficient, coupled channels type description of laser-matter interaction.
As the meaning of the individual channel functions is gauge-dependent, a finite set of channels leads to gauge-dependent results.
We argued that only in length gauge the field free ionic eigenfunctions retain their physical meaning in presence of a strong 
pulse. In contrast, in velocity gauge the same functions represent a momentum-boosted system with unphysical dynamics. 
Therefore typical physical models suggest the use of length gauge.  
This was clearly demonstrated by a mixed gauge calculation of two-electron systems, 
where the length gauge region was chosen to cover the ionic channel functions: mixed gauge calculations converge
with very few channels. Most dramatically, the single-ionization spectrum
of helium was calculated to $\lesssim10\%$ accuracy  using only the ionic ground state channel.
In contrast, in velocity gauge the single channel result is by nearly two orders of magnitude away and 
convergence could not be achieved with up to 9 channels. In pure length gauge a computation 
is out of reach because of the required discretization size.

Convergence with only the field-free neutral and very few ionic states can justify {\em a posteriori} 
wide-spread modeling of laser-atom interactions in terms of such states. It also supports the view 
that length gauge is the natural choice for this type of models. The convergence behavior 
of mixed gauge calculations --- possibly contrasted with pure velocity gauge calculations --- may
help to judge the validity of these important models in more complex few-electron systems.

\section*{Acknowledgement}
V.P.M. is a fellow of the EU Marie Curie ITN ``CORINF'', A.Z. acknowledges support by the DFG 
through the excellence cluster ``Munich Center for Advanced Photonics (MAP)'', and by the Austrian Science
Foundation project ViCoM (F41). A.S. gratefully acknowledges partial support 
by the National Science Foundation under Grant No. NSF PHY11-25915.

\input{gauge.bbl}

\end{document}

%% file: my_commands.tex
\usepackage{amsmath}
\usepackage{amssymb}
\usepackage{makeidx}
\usepackage{graphicx}
\usepackage{color}

\newcommand{\wrt}{w.r.t.\ }

\newcommand{\hide}[1]{}

\newcommand{\inv}[1]{ ^{-#1}}

\newcommand{\myvec}[1]{\vec{#1}}
\newcommand{\vr}{{\myvec{r}}}

\newcommand{\vc}{{\myvec{c}}}

\newcommand{\vp}{{\myvec{p}}}

\newcommand{\vA}{{\myvec{A}}}
\newcommand{\vB}{{\myvec{B}}}

\newcommand{\vR}{{\myvec{R}}}

\newcommand{\vunit}[1]{\hat{#1}}

\newcommand{\ur}{\vunit{r}}

\newcommand{\mymat}[1]{\widehat{#1}}

\newcommand{\mH}{\mymat{H}}

\newcommand{\mO}{\mymat{O}}

\newcommand{\mS}{\mymat{S}}

\newcommand{\cA}{\mathcal{A}} 
\newcommand{\cD}{\mathcal{D}} 

\renewcommand{\r}{\rangle}
\renewcommand{\l}{\langle}

\newcommand{\ddt}{\frac{d}{dt}}

\newcommand{\ddx}{\partial_x}
\newcommand{\ddxx}{\partial_x^2}

\newcommand{\ep}{\epsilon}

\newcommand{\la}{\lambda}

\renewcommand{\bar}{\begin{array}{ll}}
\newcommand{\ear}{\end{array}}
\newcommand{\bma}{\begin{pmatrix}}
\newcommand{\ema}{\end{pmatrix}}
\newcommand{\beq}{\begin{equation}}
\newcommand{\eeq}{\end{equation}}
\newcommand{\bel}[1]{\begin{equation}\label{eq:#1}}
\newcommand{\eel}{\end{equation}}
\newcommand{\bea}{\begin{eqnarray}}
\newcommand{\eea}{\end{eqnarray}}

\newcounter{lecture}

\newcommand{\Ef}{\mathcal{E}}
\newcommand{\vEf}{\vec{\mathcal{E}}}

\newcommand{\vna}{\vec{\nabla}}

%% file: gauge.bbl
%